# Label-free characterization of white blood cells by measuring 3D refractive index maps

Jonghee Yoon[1], Kyoohyun Kim[1], HyunJoo Park[1], Chulhee Choi[2], Seongsoo Jang[3], and YongKeun Park[1,*]

[1]*Department of Physics, Korea Advanced Institute of Science and Technology, Daejeon 305-701, South Korea*
[2]*Department of Bio and Brain Engineering, Korea Advanced Institute of Science and Technology, Daejeon 305-701, South Korea*
[3]*Department of Laboratory Medicine, University of Ulsan, College of Medicine and Asan Medical Center, Seoul 138-736, Republic of Korea.*
[*]*yk.park@kaist.ac.kr*

**Abstract:** The characterization of white blood cells (WBCs) is crucial for blood analyses and disease diagnoses. However, current standard techniques rely on cell labeling, a process which imposes significant limitations. Here we present three-dimensional (3D) optical measurements and the label-free characterization of mouse WBCs using optical diffraction tomography. 3D refractive index (RI) tomograms of individual WBCs are constructed from multiple two-dimensional quantitative phase images of samples illuminated at various angles of incidence. Measurements of the 3D RI tomogram of WBCs enable the separation of heterogeneous populations of WBCs using quantitative morphological and biochemical information. Time-lapse tomographic measurements also provide the 3D trajectory of micrometer-sized beads ingested by WBCs. These results demonstrate that optical diffraction tomography can be a useful and versatile tool for the study of WBCs.

## 1. Introduction

White blood cell (WBC) residue throughout the body plays various roles in defending the host from invaders and abnormal cells [1]. With regard to immunology, the characterization of WBCs is important for understanding the pathophysiology of many diseases, including autoimmune diseases [2], neurodegenerative diseases [3], and cancer [4]. WBCs are classified into several subsets according to their morphologies and roles in the immune system. To classify and analyze WBCs, several optical imaging methods are extensively used. For example, light microscopy with Giemsa staining, which is the standard method in clinical practice, visualizes the characteristic features of the cytoplasm and nuclei of WBCs [5, 6]. However, Giemsa staining requires chemical fixation procedures which limit live cell analysis, and only 2D images can be obtained. Confocal fluorescence microscopy enables the 3D structural images of living WBCs at a high resolution and with high molecular specificity, without physical sectioning [7, 8]. However, chemical staining procedures or genetic modifications are invasive methods, inevitably presenting significant drawbacks such as phototoxicity and photobleaching.

However, measuring the refractive indices (RIs) of biomolecules, i.e., intrinsic optical properties describing light-matter interactions, circumvents the aforementioned limitations in the study of WBCs. Quantitative phase imaging (QPI) techniques have been introduced to visualize cells and tissues, exploiting RIs as imaging contrasts [9-11]. Exploiting the principle of light interference, QPI techniques can quantitative and non-invasively measure the RI information of samples. Recently, QPI techniques have been widely applied to study the pathophysiology of various biological samples, including red blood cells [12-21], neurons [22, 23], cancer cells [24], and phytoplankton [25].

Thus far, although various efforts for the imaging WBCs using QPI techniques have been proposed [26-30], previous works have been limited to 2D imaging. Earlier research has reported morphological and biochemical changes of WBCs during WBC-mediated cytotoxicity [26, 27], pathogen infection [28, 30], and differentiation [29] by measuring the RIs of samples. Methods which use the transport of intensity equation also enable the measurement of quantitative phase information by numerical calculations, and these methods have been used to investigate WBCs [31, 32]. Although these previous approaches have demonstrated the potential of 2D QPI techniques to study WBCs, 3D tomographic measurements of WBCs have not been investigated. All previous attempts have been limited to the topographic 2D imaging of WBCs, while 3D quantitative morphological and biochemical analyses of WBCs have not been attempted. This limitation in 3D imaging is unfortunate because QPI has much to offer in research related to WBCs given its unique quantitative imaging contrast and the non-invasiveness of these processes. Furthermore, considering the complex subcellular structures of WBCs and their importance in cell classification and immunology, there is strong motivation to extend this technology to the 3D RI tomographic imaging of WBCs.

Here, we report the optical measurement of the 3D RI distributions of individual mouse lymphocytes and macrophages via optical diffraction tomography (ODT). High-resolution 3D RI tomograms of individual WBCs are constructed from multiple 2D optical phase delay images of the samples illuminated with various angles of incidence [33]. 3D RI tomogram provides quantitative morphological and biochemical information about WBCs, including the cellular dry mass, the dry mass density, volume, surface area, and sphericity. We demonstrate that 3D RI tomography enables the separation of heterogeneous populations of WBCs using quantitative structural and biochemical information. Moreover, time-lapse tomographic measurements of individual WBCs are shown to be capable of precisely visualizing the 3D trajectories of micrometer-sized beads ingested by macrophages, from which the viscoelastic properties of local cytoplasm can be investigated. The measured 3D RI tomograms of individual WBCs clearly demonstrate that ODT can be a useful and versatile tool for the study of WBCs.

## 2. Methods

2.1 Experimental setup

The optical setup is presented in Fig. 1(a). Complex optical fields of a sample, containing both the amplitude and phase images, are recorded using a Mach-Zehnder interferometric microscope for various illumination angles [34, 35]. A laser beam from a diode-pumped solid state laser ($\lambda$ = 532 nm, 100 mW, Shanghai Dream Laser Co., Shanghai, China) is split into two arms by a beam splitter (BS1). One arm is used as a reference beam, and the other arm is tilted by a dual-axis scanning galvanometer (GVS012, Thorlabs, Newton, NJ, USA) for varying the angle of the illumination beam impinging to the sample.

Hologram of the sample is generated by interference of two beams, which is recorded by a high-speed CMOS camera (1024 PCI, Photron USA Inc., San Diego, CA, USA) with a frame rate of 1,000 Hz. Typically, 300 holograms of the sample, illuminated by plane waves with various illumination angles (-70° to 70° at the sample plane), are recorded for reconstructing one RI tomogram. Details about the experimental setup used to measure complex optical fields can be found in the literature [33].

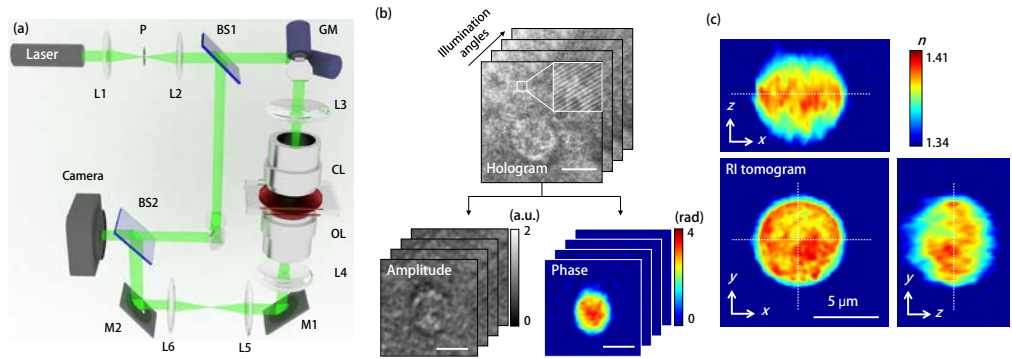

Fig. 1. Schematic of the experimental ODT setup and the procedure of a 3D RI tomogram reconstruction. (a) A Mach-Zehnder interferometric microscope equipped with a 2D scanning galvanometer-based mirror. BS1–2, beam splitters; GM, galvano mirror; OL, objective lens; CL, condenser lens; M1–2, mirrors; P, pinhole; L1–6, lenses. (b) Holograms are recorded with various illumination angles (top) and the retrieved amplitude and the phase images corresponding to a hologram at a specific illumination angle (bottom). Inset: zoomed-in view of spatially modulated interference patterns. Scale bar, 5 μm. (c) Cross-sectional slices of a RI tomogram of a WBC. Scale bar, 5 μm.

2.2 Tomogram reconstruction

From the measured multiple 2D complex amplitude images of a sample, a 3D RI tomogram of individual samples is reconstructed via the ODT algorithm (Fig. 1(b)), which is analogous to 3D computed tomography in X-ray. First, the complex optical fields are extracted from measured holograms using a field retrieval algorithm [36]. Multiple complex amplitude images obtained with various illumination angles are 2D Fourier transformed. Then the spectral information are mapped onto a surface, so-called Ewald sphere, in 3D Fourier space. Finally, 3D RI tomogram is reconstructed by applying 3D inverse Fourier transformation to the mapped 3D Fourier space. Due to the limited numerical aperture (NA) of the used imaging system, there exist missing spectral information. To fill this missing information, Gerchberg-Papoulis algorithm based on a non-negativity constraint was used [37]. The theoretical lateral and axial resolution of the reconstructed tomogram is 111 and 354 nm, respectively, which was calculated from the maximum range of the Fourier spectra [34]. The lateral and axial resolution was experimentally measured as 373 and 496 nm, respectively, by analyzing the edge of the reconstructed tomograms of polystyrene beads. The detailed reconstruction process including a MatLab™ code can be found in our previous work [33].

2.3 Cell preparation

All experiments used 7- to 10-week-old male Balb/c mice (Orient Bio Inc., Gapyeong, Korea). Lymphocytes and macrophages were collected from mice peripheral blood and peritoneal cavity, respectively. Peripheral blood obtained from the heart of euthanized mice was added to heparin (10 U/ml). Heparinized blood was diluted with an equal volume of phosphate-buffered saline (PBS, Welgene Inc., Gyeongsan, Korea) and layered on 3 ml of lymphocyte separation medium (MP Biomedicals, Irvine, CA, USA) in a 15 ml conical tube. This solution was centrifuged at 400 g at room temperature for 20 min to separate lymphocytes from red blood cells. Lymphocytes layer was collected and washed 2 times with PBS. The cells were resuspended in Dulbecco's modified Eagles' medium (DMEM, Gibco, Big Cabin, OK, USA) supplemented with 10% heat-inactivated fetal bovine serum (FBS). To isolate macrophages, ice cold PBS (with 3% FBS) was injected into the peritoneal cavity of euthanized mice. Injected fluid was collected from peritoneal cavity and centrifuged at 250 g for 8 min. The cells were maintained in DMEM for 2 days before experiment [38]. Macrophage and lymphocyte were sandwiched between two cover slips and imaged at room temperature. For phagocytosis analysis, 1 μm polystyrene beads (89904, Sigma-Aldrich, St. Louis, USA) were added to culture medium of cultured macrophages. Measurements were performed 1 hour after beads addition. The sample preparation procedures and the methods were approved by the Institutional Review Board (KA-2015-03).

## 3. Results and discussions

### 3.1 Label-free 3D RI imaging of lymphocytes and macrophages

We initially measured the 3D RI tomograms of individual lymphocytes and macrophages (See Methods). Figures 2(a, c) present cross-sectional images of 3D RI tomograms on various axial planes: 1 μm below the focal plane (left column), the focal plane (center column), and 1 μm above the focal plane (right column) along the axial axis. The RI distribution map clearly exhibits the plasma membrane shapes and intracellular structures of a lymphocyte and a macrophage.

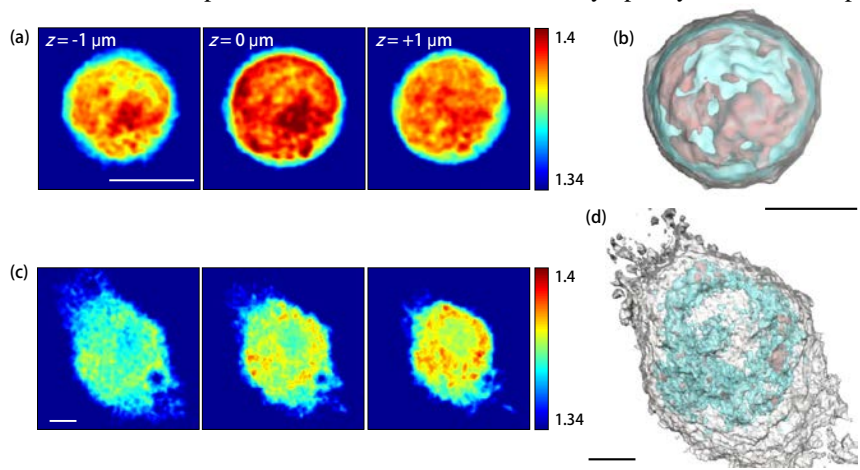

Fig. 2. 3D visualization of WBCs. Cross-sectional slices of the RI distribution of (a) a lymphocyte and (c) a macrophage at various axial planes. Scale bar, 4 μm. Rendered iso-surfaces of the RI map of (b) a lymphocyte and (d) a macrophage. For visualization purposes, the threshold iso-surfaces of the plasma membrane (white) and the internal structures (cyan, red) are set to 1.345, 1.375, and 1.39, respectively. Scale bar, 3 μm.

The lymphocyte shows a round cell morphology, small cytoplasmic volume, and one large nucleus with high RI values (Fig. 2(a)). Although the macrophage contains one nucleus, it shows a larger cytosolic volume compared to the lymphocyte as well as multiple vacuoles in the cytoplasm (Fig. 2(b)). 3D rendered iso-surfaces of the RI values of a lymphocyte and a macrophage respectively visualize 3D morphology of plasma membrane (white) and intracellular organelles (cyan and red), depending on RI values (Figs. 2(b), (d)). For visualization purposes, the threshold iso-surfaces of the plasma membrane (white) and the internal structures (cyan, red) are set to 1.345, 1.375, and 1.39, respectively. The structural information of the 3D RI tomogram are comparable to images obtained by confocal fluorescence microscopy [7] or electron microscopy [39].

### 3.2 Quantitative characterizations of WBCs using 3D RI tomograms

Measurements of the 3D RI distribution of WBCs can provide morphological (cellular volume, surface area, and sphericity) and biochemical (dry mass and dry mass density) information about individual WBCs. To demonstrate the quantitative imaging capability, we measured the 3D RI tomograms of lymphocytes and macrophages and retrieved morphological and biochemical information pertaining to individual WBCs. Despite the fact that lymphocytes and macrophages are subtypes of a mononuclear cell, they have distinct morphologies and various roles in the immune system. These results are presented in Fig. 3.

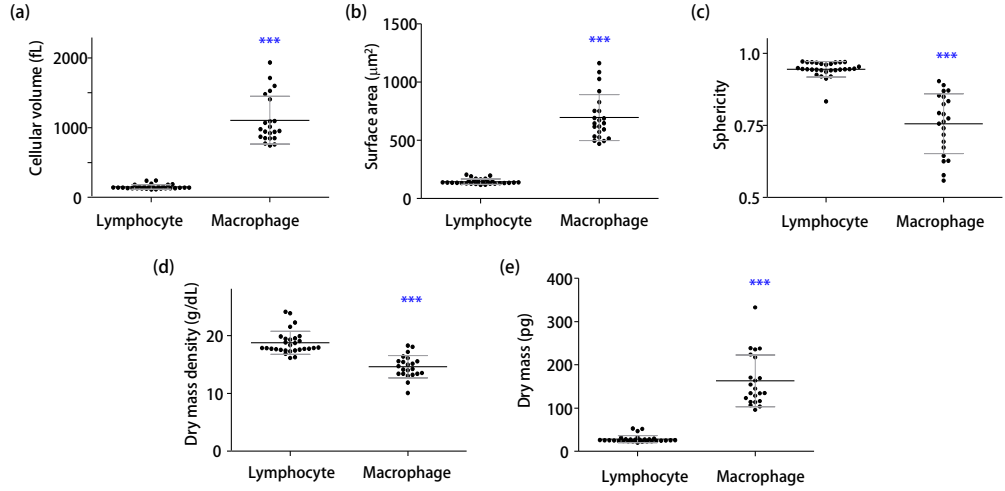

Fig. 3. Quantified morphological and biochemical information about individual lymphocytes ($n = 29$) and macrophages ($n = 22$). (a) Cellular volume, (b) surface area, (c) sphericity index, (d) dry mass density, and (e) dry mass. Each symbol represents an individual cell measurement and the horizontal black line indicates the mean value with the vertical line of standard deviation. The symbol *** indicates a $p$-value of $< 0.001$ in the comparison of the lymphocytes and macrophages according to Student's $t$ test.

In order to obtain the morphological information, the iso-surfaces retrieved from the measured 3D RI tomograms are used. The cellular volume $V$ was obtained by integrating voxels which have RI values higher than that of surrounding media in the measured 3D RI tomograms. The mean values of the cellular volumes of individual WBCs are $150.85 \pm 31.68$ fL ($n = 29$) and $1106.08 \pm 334.28$ fL ($n = 22$), for lymphocytes and macrophages, respectively (Fig. 3(a)). The cell surface area $S$ was obtained by calculating the surface area of the RI iso-surface corresponding to the cell boundaries. The mean values of the cellular surfaces areas are $144.76 \pm 21.88$ μm² and $695.84 \pm 192.91$ μm² for lymphocytes and macrophages, respectively (Fig. 3(b)). From the measured S and V values, the sphericity SI can be directly calculated, as follows:

$$SI = \pi^{1/3}(6V)^{2/3}/S. \qquad (1)$$

The SI provides a dimensionless measure of how spherical an object is; The SI of a perfect sphere is 1 and the SI of a plane is 0. The calculated mean values of the SI are determined to be $0.94 \pm 0.03$ and $0.76 \pm 0.10$ for lymphocytes and macrophages, respectively (Fig. 3(c)).

Measurements of 3D RI tomograms provide quantitative information about local concentrations of non-aqueous biomolecules such as intracellular proteins inside cells. From the average RI values in the cytoplasmic areas of WBCs, the mean dry mass density $\rho$ was calculated. The RI value of the cytoplasm is proportional to the concentration of non-aqueous molecules in the cytoplasm (mostly proteins) according to the relationship $n = n_0 + \alpha\rho$ [40], where $n$ is the refractive index of protein, $n_0$ is the refractive index of the surrounding medium, and $\alpha$ is the specific refractive index increment (RII) of a protein species. Most proteins have similar RII values [41, 42]; thus, we used the specific RII value of 0.2 mL/g for the calculations in this study. The total cellular dry mass $m$ of a WBC [43-45] was obtained by integrating the local dry mass density $\rho$ over the cell volume $V$. The mean values of the dry mass density are $18.78 \pm 1.97$ g/dL and $14.64 \pm 1.90$ g/dL, and the mean values of the dry mass are $28.50 \pm 8.05$ pg and $163.16 \pm 58.75$ pg for lymphocytes and macrophages, respectively (Figs. 3(d), (e)). The majority of the measured lymphocytes were round shapes whereas the macrophages displayed various non-spherical morphologies. The cytoplasm of a lymphocyte is mainly composed of one large nucleus with high RI values, but not in a macrophage. Thus, the cellular dry mass densities of macrophages are significantly lower than those of lymphocytes (Fig. 3(d)). Although the cellular dry mass densities are low in macrophages, the total dry masses of macrophages are approximately four times higher than those of lymphocytes due to their greater cell volumes (Figs. 3(d), (e)).

3.3 Heterogeneous populations even in isolated lymphocytes

Heterogeneous populations in isolated lymphocytes were observed in the measured 3D RI tomograms of WBCs. Figures 4(a, b) present quantitative analyses of representative small and large lymphocyte, which results are consistent with previous reports [46, 47]. Previous studies report that lymphocytes can be distinguished into small and large lymphocytes, whereas previous approaches only considered the 2D mean diameters $D$ of cells obtained with light microscopes [46, 47]. The values of the 2D mean diameters obtained in this study are 6.29±0.27 µm and 8.33±0.97 µm for small and large lymphocytes, respectively. These 2D mean diameters are in good agreement with those in previous reports [46, 47].

In order to demonstrate the capability of the present approach, we retrieved morphological and biochemical information pertaining to both small and large lymphocytes (Figs. 5(b)-(f)). The cellular volume and surface area of the large lymphocytes are 2.10 times larger and 1.62 times larger than those of the small lymphocytes, respectively (Figs. 5(b), (c)). However, both cases demonstrate similar spherical shapes and similar RI values.

For a further demonstration of the applicability of the present method, we investigated morphological alternation in lymphocytes due to immunogenic stimulation. In order to stimulate lymphocytes, we used lipopolysaccharide (LPS, L3012, Sigma-Aldrich, St. Louis, USA), which the most abundant component in bacterial cell walls. It is known that a treatment of LPS induces functional and morphological changes in immune cells [48, 49]. Isolated lymphocytes were stimulated by a LPS treatment (10 ng/mL) for three hours before measurements were taken. These results are presented in Fig. 4(c) and Fig. 5. The LPS-treated lymphocytes exhibit significant alterations in their morphologies Sphere-like lymphocytes are flattened (the mean SI value decreases from 0.94±0.03 to 0.92±0.05), and the formation of multiple filopodia is clearly shown. We observed that the cellular volume and dry mass of lymphocytes increase upon a LPS treatment, which is consistent with the findings in previous reports [50, 51].

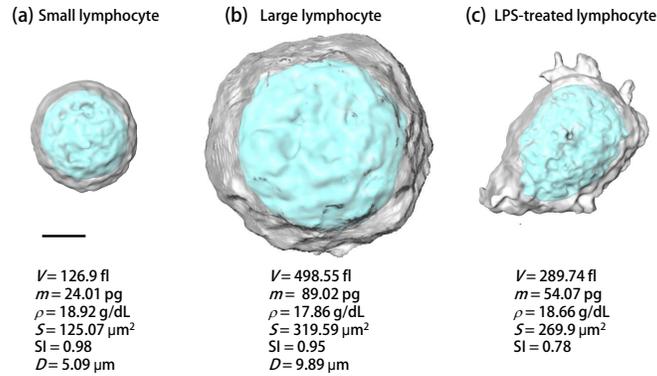

Fig. 4. Quantitative analyses of representative lymphocytes in heterogeneous populations. 3D rendered RI surfaces of (a) a small lymphocyte, (b) a large lymphocyte, and (c) a LPS-treated lymphocytes. The cellular volume $V$, cellular dry mass $m$, dry mass density $\rho$, surface area $S$, sphericity $SI$, and 2D mean diameter $D$ are specified. Scale bar, 2 µm.

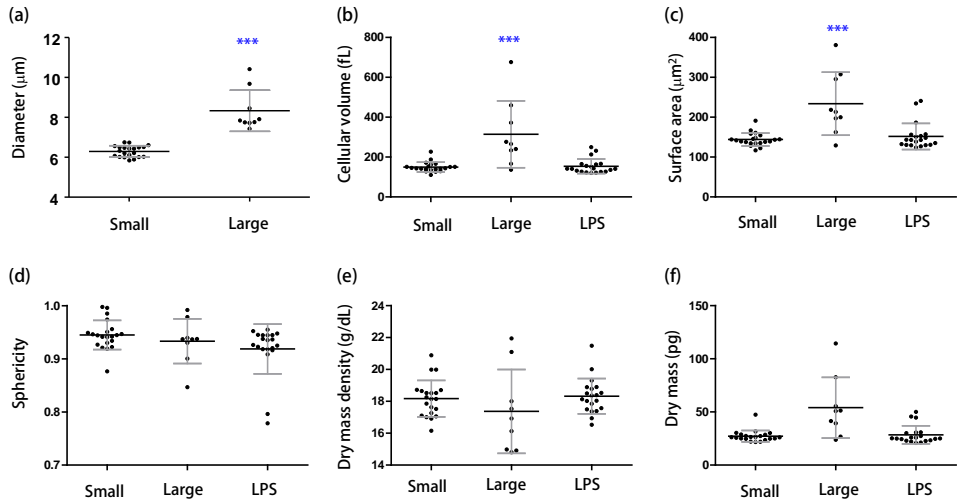

Fig. 5. Heterogeneous populations in isolated lymphocytes. Quantified morphological and biochemical information about small (n=21), large (n=9), and LPS-treated lymphocytes (n=22). ($n$ = 29) and macrophages ($n$ = 22). (a) Diameter, (b) cellular volume, (c) surface area, (d) sphericity, (e) dry mass density, and (f) dry mass. Each symbol represents an individual cell measurement and the horizontal black line indicates the mean value with the vertical line of standard deviation. The symbol *** indicates a $p$-value of < 0.001 in the comparison of the lymphocytes and macrophages according to Student's $t$ test.

## 3.4 3D dynamics of micrometer-sized beads ingested by a macrophage

To demonstrate the dynamic 3D imaging capability, we emulated the phagocytosis of macrophages using micrometer-sized beads and measured time-lapse dynamic 3D RI tomograms of macrophages with engulfed microspheres. Phagocytosis is a crucial mechanism in the immune system; macrophages identify, track, and ingest harmful pathogens or abnormal cells, eliminating them to protect the host [52, 53]. We added 1 μm of polystyrene beads to a culture medium of macrophages for 1 hour before the measurement (see Methods). Figure 6(a) clearly displays three ingested beads inside a macrophage, as indicated by the RI value of polystyrene (approx. 1.41) inside the cytoplasm.

In order to study the temporal dynamics of the ingested beads within a macrophage, a series of 3D RI tomograms was obtained every 10 s. All three microspheres exhibit movement on the *x-y* plane and in the axial direction (Movie 1). From the series of 3D RI tomograms, the 3D trajectories of each of the beads were retrieved, as presented in Fig. 6(b). The tracking of the microspheres inside living cells provides valuable information about the viscoelastic properties of the cell cytoplasm. Most previous approaches used 2D projected trajectories of target objects in order to assess micro-rheological properties of cells. However, the 3D tracking of objects is technically challenging; holographic field measurements, 3D confocal microscopy, or multiple projection processes with various views have been exploited. Herein, we demonstrate that the dynamic 3D tomographic measurements of cells with engulfed microspheres readily enables an investigation of the viscoelastic properties of the cytoplasm of cells while also visualizing the overall cell shape. This is possible because the quantitative and non-invasive capabilities of the present method, i.e., engulfed microspheres with a specific RI, can be readily identified in measured 3D RI tomograms.

The movements of ingested beads may appear to be random Brownian motion, but particle movements within living cells show active non-equilibrium dynamics [54, 55]. To the analyze movements of engulfed particles within a macrophage further, we calculated the mean square displacements (MSDs). A double-logarithmic plot of the MSDs of three beads as a function of the lag time $\tau$ is shown in Fig. 6(c), with fitted slopes for both the fast dynamics ($\tau = 10-40$ s) and slow dynamics ($\tau = 50-100$ s). In the fast dynamics, or short timescales, all of the beads show slopes smaller than 1.0, indicating sub-diffusive motion. However, one of the beads exhibits super-diffusive motion (slope = 1.49) in the slow dynamics, suggesting active transport of the bead within the macrophage

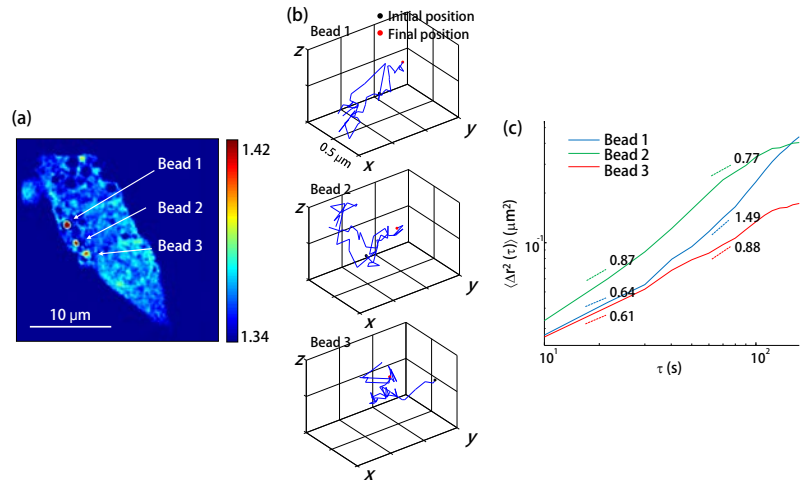

Fig. 6. Time-lapse measurement of the 3D RI tomograms and 3D trajectories of polystyrene beads engulfed by a macrophage. (a) Cross-sectional slice of a 3D RI tomogram of a macrophage on the focal plane. Three engulfed beads are indicated. (b) 3D trajectories of polystyrene beads in (a). The initial and final positions of each bead are indicated by the black and red circles, respectively. (c) MSD of the beads in (b) as a function of the lag time. Dashed lines indicate the fitted slope.

## 4. Conclusion

Here, we present the label-free quantitative RI tomographic imaging of WBCs and the characterization of individual WBCs. We used an interferometric microscope based on Mach-Zehnder interferometry equipped with a 2D galvano mirror. Using this setup, 3D RI maps of individual WBCs are constructed from multiple 2D optical field images obtained with various illumination angles. We demonstrate that 3D RI tomograms of individual WBCs provide quantitative information about the structural and biochemical characteristics of WBCs, including the cellular volumes, surface areas, sphericities, dry mass densities, and total dry mass. In addition, the present approach was applied to an investigation of subpopulations of lymphocytes and to morphological alternations in LPS-treated lymphocytes at the individual cell level. Furthermore, we found that dynamic measurements of 3D RI tomograms of WBCs can be utilized for the study of phagocytosis and of the viscoelasticity of WBSs using a model of engulfed microspheres. The 3D trajectories of engulfed polystyrene microspheres were obtained from dynamic 3D RI tomograms, and these dynamics were quantified by calculating the MSD.

Because RI is an intrinsic optical parameter of a material, the present approach does not require the use of exogenous labeling agents, staining procedures, or genetic modifications. This label-free capability of RI measurements can offer long-term observation of individual cells without phototoxicity and photobleaching processes, which can perturb certain cellular functions. More importantly, this RI, an intrinsic optical contrast, was measured quantitatively according to the principles of interferometry or holography. We demonstrated the potentials of the quantitative imaging capability by investigating the quantitative characteristics of the measured 3D RI tomograms.

We undertook dynamic measurements of 3D RI tomograms at a speed of 0.1 Hz. However, the acquisition, reconstruction, and visualization speed can be enhanced significantly using a graphics processor unit (GPU) and a sparse sampling method [56]. Recently, our group successfully measured 3D RI tomograms at 60 Hz [57]. From a technical perspective, the present approach can be expanded by combining it with other imaging modalities, including fluorescence imaging [26] and light polarization [58-60]. Furthermore, current optical bright-field microscopes can be converted for ODT purposes using a recently developed QPI unit [61, 62], which will expand the applicability of the present method in immunology without the need for complicated optical alignments. One of the limitations of the present approach is the limited molecular specificity. Although the molecular specificity of label-free RI imaging may not be as high as its fluorescence-labeled counterparts, spectroscopic RI measurements may be used to separate molecules spectrally with various optical dispersions [63-67].

In sum, we envision that ODT will be a useful tool in the study of immunology given its ability to investigate the precise morphologies of cells and subcellular organelles, including nucleoli, and to provide detailed biochemical information. The present approach can be readily applied to the study of WBCs in various disease conditions. For example, leukemia results in very high number of WBCs and induces abnormalities in the morphologies and functions of WBCs [68, 69]. 3D RI tomograms of individual WBCs can be utilized to detect abnormalities in WBCs and to diagnose leukemia in the early stage of the disease.

**Acknowledgments**

This work was supported by KAIST, National Research Foundation (NRF) of Korea (2012R1A1A1009082, 2013K1A3A1A09076135, M3C1A1-048860, 2013M3C1A3063046, NRF-2012-M3C1A1-048860, 2013R1A1A3011886, 2014M3C1A3052537), and KUSTAR-KAIST project. KHK is supported by Global Ph.D. Fellowship from NRF.